\documentclass[12pt]{article}
\usepackage{amsmath}
\usepackage{amssymb}
\usepackage{graphicx}

\begin{document}
\baselineskip=18pt
\begin{titlepage}
\begin{flushright}
{\small KYUSHU-HET-100}\\[-1mm] hep-ph/0610402%
\end{flushright}
\begin{center}
\vspace*{11mm}

{\large\bf%
Low-energy Variety of Asymmetric SUSY Flavor Structure%
}\vspace*{8mm}

Kenzo Inoue, Kentaro Kojima, and Koichi Yoshioka%
\vspace*{1mm}

{\it Department of Physics, Kyushu University, Fukuoka 812-8581, Japan}

\vspace*{3mm}

{\small (October, 2006)}
\end{center}
\vspace*{5mm}

\begin{abstract}\noindent%
In this letter we study the low-energy phenomenology and cosmological
implications of supersymmetric grand unified theory with 
asymmetric flavor structure,
which is suggested by the recent observation of
fermion masses and mixing angles. The predictions of the scenario are
rather dependent on a Yukawa parameter fixed by group-theoretical
argument. A reduced value of the parameter gives a resolution to the
sign problem of supersymmetric Higgs mass $\mu$, with which the theory
becomes simultaneously consistent with the experimental data of
bottom/tau mass ratio, flavor-changing rare decay of bottom quark, and
the muon anomalous magnetic moment. The relic abundance of the lightest
superparticle as cold dark matter of the universe is also investigated
in light of the three-years WMAP result.
A new source of flavor violation is found
in the D-term induced scalar masses, that is a distinctive signature
of the generation asymmetry.
\end{abstract}
\end{titlepage}

\newpage

The minimal supersymmetric standard model is conceived to be
one of the promising theoretical candidates beyond the standard model
for its brevity and attractive features. A remarkable property of 
supersymmetric standard models is the unification of gauge coupling
constants at some high-energy scale~\cite{unify}. This motivates us to
promote it into grand unified theory (GUT)~\cite{GG} and various
attempts have been devoted to constructing realistic models and
understanding their phenomenological implications.

The quarks and leptons are generally unified into multiplets of
GUT gauge group. That seems to conflict with disparate flavor
structures, namely the observed fermion masses and mixing angles. For
example, the well-known GUT relation between down-quark and charge-lepton
Yukawa couplings leads to the same mass eigenvalues of quarks and
leptons, which is successful only for the third generation. Further if
each one-generation matter is combined into a single multiplet, the
quark and lepton mixing matrices have similar forms, inconsistent with
the experimental observation. Therefore a naive unification
hypothesis should be modified in the flavor (Yukawa coupling) sector.

Lopsided flavor structure in the generation space is one of 
promising ways to remedy the flavor difficulties in grand
unification. It has recently been shown, for example, that the flavor
asymmetry is dynamically realized by introducing matter multiplets in
a generation-dependent fashion. With such a experimentally suggested
form of flavor couplings, novel behaviors emerge in the electroweak
gauge symmetry breaking and the mass spectrum of
superparticles. Extending the previous study~\cite{nuEWSB}, in this
letter we further discuss phenomenological and cosmological issues
such as flavor-violating rare processes and the relic dark matter
abundance in the universe.

\bigskip\bigskip

In the analysis of this letter, we assume the following general form
of Yukawa couplings at the GUT scale, where the electroweak gauge
couplings meet:
\begin{equation}
  Y_u \,=\,Y_\nu \,=\,
  \left(\begin{array}{ccc}
    ~~ & ~~ & \\
    & & \\
    & & y_t
  \end{array}\right), \qquad 
  Y_d \,=\,
  \left(\begin{array}{ccc}
    ~~~ & & \\
    & & x_d\,y' \\
    & & y
  \end{array}\right), \qquad
  Y_e \,=\,
  \left(\begin{array}{ccc}
    ~~ & ~ & \\
    & & \\
    & \,y' & y
  \end{array}\right),
  \label{Y}
\end{equation}
at the leading order. The blank entries take negligibly small
values. The top quark and tau neutrino Yukawa coupling $y_t$ come from
the minimal Higgs coupling. On the other hand, the down-quark and
charged-lepton Yukawa matrices, $Y_d$ and $Y_e$, have flavor
asymmetrical forms. The latter structure of Yukawa couplings is
incorporated in grand unified framework~\cite{GUT} and more reasonably
is achieved dynamically by extra matter (and also Higgs)
multiplets~\cite{ABB,SNY,BKY}. For example, in the $SO(10)$ language,
the third-generation fermions and the up-type Higgs field come from 16
and 10-plets, respectively. 
The second generation fermions are generally given by some mixtures of 16 and
10-plets. Consequently the down-type Higgs is a mixture of the 
above $10_H$-plets and other pieces: 
$H_d\subset10_H\cos\theta+({\rm other})\sin\theta$. 
This leads to a GUT relation  $y=y_t\cos\theta$. The
Yukawa couplings (\ref{Y}) solve the aforementioned flavor mixing
problem in unified theory; the observed large lepton mixing is
explained if $y\simeq y'$, while this does not disturb the small
(left-handed) quark mixing.

Another important quantity is $x_d$: the ratio 
of $(Y_d)_{23}$ and $(Y_e)_{32}$. A deviation from the naive unified
relation $x_d=1$ solves the mass eigenvalue problem of down quarks and
charge leptons in unified theory. The parameter $x_d$ is controlled by
group-theoretical factors. Two simple limits 
are $x_d=1$ and $x_d=-1/3$. The former coupling is seen as the limit
of Yukawa unification. The latter one, as a renormalizable coupling,
originates from a higher-rank Higgs field and is suitable for
reproducing the second (and first) generation fermion
masses~\cite{GJ}. These two limits respectively well-capture the
distinct phenomenology in the cases that $|x_d|$ is large
($\gtrsim0.5$) and small ($\lesssim0.5$). While $x_d$ could generally
take a discrete and predictive value, we consider in this letter these
limits as reasonable examples, and call $x_d=1$ as the aligned
bottom/tau case and $x_d=-1/3$ as the reduced bottom/tau case.

Without specifying the details of GUT model, we consider the twist 
parameter $\theta$ as free and the ratio $x_d$ as its typical 
values $x_d=1$ and $-1/3$. The other couplings $y$ and $y'$ are
responsible for the top quark mass and the large mixing of atmospheric
neutrinos, and therefore experimentally determined. In the following
analysis, the two parameters $\theta$ and $x_d$ play important roles
to clarify the phenomenological and cosmological implications of
flavor asymmetry. 

\bigskip\bigskip

Let us start by discussing about the third-generation fermion masses
from the Yukawa matrices (\ref{Y}). The mass prediction at low-energy
region depends on the couplings $\theta$ and $x_d$, and are also
affected by supersymmetry-breaking parameters through low-energy
threshold corrections at the decoupling of superparticles. The latter
effect is known to provide an important suggestion for
phenomenologically viable classes of superparticle mass spectrum.

\begin{figure}[t]
\begin{center}
\includegraphics[width=13cm]{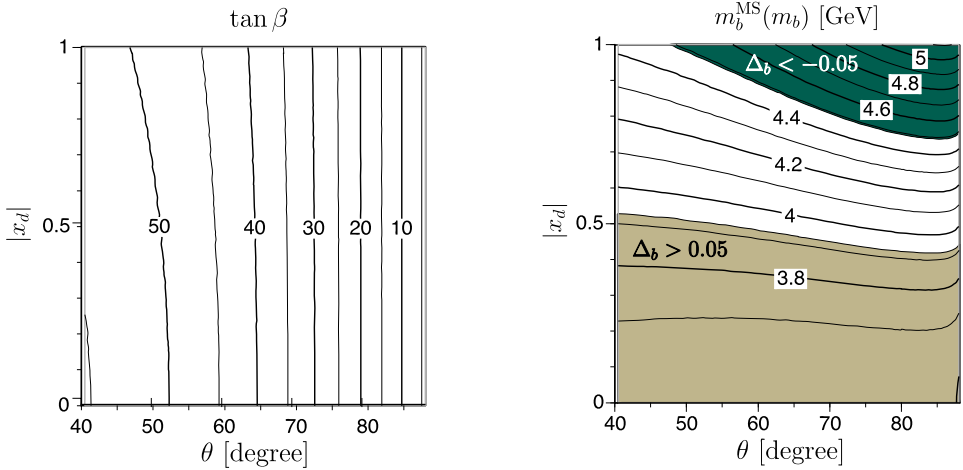}
\end{center}
\caption{The predictions of bottom quark mass and $\tan\beta$. The
two-loop renormalization group evolution down to the electroweak scale
is solved. The low-energy threshold corrections are fixed as typical
values for the top quark and tau lepton 
masses ($\Delta_t=0.05$ and $\Delta_\tau=-0.02$). In the yellow
(green) shaded region, the correction $\Delta_b$ needs to be larger
than 0.05 (smaller than $-0.05$) to have the correct mass
eigenvalue.\bigskip}
\label{fig:mbtanb}
\end{figure}

We now evaluate the bottom quark mass and $\tan\beta$, the ratio of
the two electroweak Higgs doublets, from the observed values of top
quark mass $m_t^{\rm pole}=172.7$~GeV~\cite{CDF} and tau lepton 
mass $m_\tau^{\rm pole}=1777$~MeV~\cite{PDG}. The right-handed
neutrinos are assumed to decouple at $10^{14}$~GeV and induce light
neutrino Majorana masses. (For the other details of the evaluation
procedure, see Ref.~\cite{nuEWSB}.) \ In Fig.~\ref{fig:mbtanb}, the
predictions of bottom quark mass and $\tan\beta$ are displayed as the
functions of $\theta$ and $x_d$.
The analysis does not include the parameter 
region $\theta\lesssim40^\circ$ and $|x_d|\gtrsim1$. This is because
in the former region the electroweak breaking vacuum is unstable and
the latter region is less preferable for the strange quark and muon
masses. It is found from the left figure that the twist 
parameter $\theta$ controls a relative strength between top and bottom
Yukawa couplings, thereby the value of $\tan\beta$, whereas another
parameter $x_d$ is almost irrelevant to $\tan\beta$. 

The bottom quark mass prediction shown in the right figure does not
include the low-energy threshold correction $\Delta_b$ to the bottom
quark mass. (The threshold corrections $\Delta_i$ is defined 
by $y_i'=y_i(1+\Delta_i)$ and evaluated at the decoupling scale of
superparticles.) \ It can be therefore extracted from the figure what
amount of $\Delta_b$ is required, depending mainly on the 
parameter $x_d$. Generally speaking, a smaller value of $|x_d|$ leads
to a smaller bottom/tau mass ratio and in turn needs a constructive
contribution $\Delta_b>0$. This is why we call $|x_d|\ll1$ as the
reduced bottom/tau case. In the yellow (green) shaded region in
Fig.~\ref{fig:mbtanb}, the threshold correction should 
satisfy $\Delta_b>0.05$ ($\Delta_b<-0.05$) in order to obtain the
experimental range $m_b(m_b)=4.2\pm 0.07$~GeV in
the $\overline{\rm MS}$ scheme~\cite{PDG}. On the other hand, 
the $x_d=1$ case leads to equal tree-level masses of bottom quark and
tau lepton at the GUT scale, and is referred to as the aligned
bottom/tau case throughout this letter. For a not so 
small $\tan\beta$, the correction $\Delta_b$ is dominated by finite
pieces~\cite{deltab} and not decoupled. It has the following property;
(i) the sign of correction is equivalent to that of the supersymmetric
Higgs mass parameter $\mu$, and (ii) the size of correction is
suppressed if the superparticle mass spectrum is hierarchical, 
i.e.\ the scalar quarks are much heavier than the gauginos and
higgsinos. These features lead to a general, important implication for
superparticle mass parameters in the present model. 
For the reduced bottom/tau case, a non-vanishing $\Delta_b>0$ is
needed, from which fact one expects that the $\mu$ parameter is
positive and superparticles tend to be degenerate in mass. On the
other hand, the aligned bottom/tau case makes the situation opposite,
resulting in a negative $\mu$ and hierarchical superparticle spectrum,
i.e.\ relatively heavy scalar quarks. The mass prediction of
superparticles is thus found to have a correlation with the 
parameter $x_d$ via threshold corrections to the bottom quark mass. It
is mentioned that there may be a possibility that small corrections
are supplied by a cancellation among various diagrams, but we do not
consider such a model-dependent case in this letter.

\bigskip\bigskip

The next issue is to analyze the superparticle mass spectrum in order
to realize the radiative electroweak symmetry breaking
(EWSB)~\cite{EWSB} in addition to the third-generation fermion masses
discussed above. The property of the EWSB vacuum depends on the
parameters $\theta$ and $x_d$ via radiative corrections. At the GUT
scale, the standard gaugino masses are $M_{1/2}$ and all scalar
trilinear couplings are set to $A_0$. The Higgs mixing mass parameters
are determined by solving the EWSB conditions at the electroweak
scale. We assume there are no flavor off-diagonal elements in
supersymmetry-breaking parameters at the boundary scale. They are
therefore given by the following form (irrespectively of $x_d$):
\begin{eqnarray}
  m_{\tilde Q_{ij}}^2 \!&=& m_{\tilde u_{ij}}^2 \;=\;
  m_{\tilde e_{ij}}^2 \;=\; (m_0^2-D)\,\delta_{ij}, \\
  m_{\tilde d_{11}}^2 \!&=& m_{\tilde L_{11}}^2 \;=\; 
  m_{\tilde d_{33}}^2 \;=\; m_{\tilde L_{33}}^2 \;=\; m_0^2+3D, 
  \label{dl13} \\
  m_{\tilde d_{22}}^2 \!&=& m_{\tilde L_{22}}^2 \;=\; m_0^2-2D,
  \label{dl2} \\
  m_{\tilde \nu_{ij}}^2 \!&=& (m_0^2-5D)\,\delta_{ij}, \\
  m_{H_u}^2 \!&=& m_0^2+2D, \\
  m_{H_d}^2 \!&=& m_0^2+(-2\cos^2\theta+3\sin^2\theta)D,
\end{eqnarray}
where $m_{\tilde Q}^2$, $m_{\tilde u}^2$, $m_{\tilde e}^2$, 
$m_{\tilde d}^2$, $m_{\tilde L}^2$, $m_{\tilde \nu}^2$, $m_{H_u}^2$ and
$m_{H_d}^2$ are the scalar masses for quark doublets, up-type quarks,
charged leptons, down-type quarks, lepton doublets, right-handed
neutrinos, up-type Higgs and down-type Higgs, respectively. We
generally include the D-term contribution associated with the
breakdown of unified gauge symmetry [e.g.\ $SO(10)$ in the above
example]. It is noticed that, while the D-term contribution comes from
the gauge interaction, its effects appear in a flavor-dependent way 
in $m_{\tilde d}^2$ and $m_{\tilde e}^2$. The down-type Higgs mass
also deviates from the universality and depends on $\theta$. These
facts are important consequences of the generation twisting, 
and would induce sizable flavor-violating effects, as
will be discussed.

Given the boundary conditions of coupling constants, we solve the
renormalization group evolution from the GUT scale to the electroweak
one. The requirements for the EWSB vacuum are then expressed
approximately as the predictions of physical particle masses:
\begin{eqnarray}
  |\mu|^2 &\simeq& -m_{H_u}^2-\frac{M_Z^2}{2}, \\[1mm]
  M_A^2 &\simeq& m_{H_d}^2-m_{H_u}^2-M_Z^2,
\end{eqnarray}
where $M_Z$ is the $Z$ boson mass. The quantity $\mu$ is related to
the higgsino masses, and $M_A^2=m_{H_u}^2+m_{H_d}^2+2|\mu|^2$ is the
CP-odd neutral Higgs mass. The former is given by the up-type Higgs
soft mass and the latter one by the difference between up and
down-type Higgs soft masses. A given set of the Yukawa 
parameters $\theta$ and $x_d$, the EWSB predictions are derived from
the mass parameters at the GUT scale. For example,
\begin{eqnarray}
  |\mu|^2 &=& +0.06m_0^2+1.42M_{1/2}^2+0.08A_0^2-0.24A_0M_{1/2}
  -1.96D-\frac{M_Z^2}{2},  \label{mu} \\
  M_A^2 &=& -0.11m_0^2+0.12M_{1/2}^2-0.00A_0^2-0.02A_0M_{1/2}
  -1.43D-M_Z^2,
\end{eqnarray}
for $\theta=45^\circ$ and $x_d=1$, corresponding 
to $\tan\beta\simeq51$. The coupling dependences of the numerical
coefficients are found in~\cite{nuEWSB}.

In fact, the formula of the higgsino mass $|\mu|$ is almost
insensitive to $\theta$ and $x_d$ because the EWSB prediction 
of $|\mu|$ is correlated only to the up-type Higgs mass and not
directly connected to the down-quark and charged-lepton sectors at
one-loop level. It is found from eq.~(\ref{mu}) that a negative
contribution, which make $|\mu|$ small, is obtained only through a
positive D term. Such a suppressed value of $\mu$ parameter is
obtained in the large $|x_d|$ case for the bottom quark mass being
within the experimentally allowed range. Otherwise, the higgsino mass
scale becomes larger than the gaugino mass scale, and the threshold
correction $\Delta_b$ is not generally unsuppressed.

Another EWSB prediction of $M_A$ has a particular response to the
parameters $\theta$ and $x_d$. In the reduced bottom/tau case (with
small values of $|x_d|$ and $\cos\theta$) decreases the effects of
down-quark and charged-lepton Yukawa couplings in the
renormalization-group evolution of down-type Higgs 
mass $m_{H_d}^2$. That makes $m_{H_d}^2$ larger in the infrared regime
and the CP-odd neutral Higgs mass is raised. In the above formula
of $M_A^2$, the coefficient of $m_0^2$ therefore increases 
as $\theta$ and changes its sign from negative to positive 
around $\theta\simeq50^\circ$ depending on $x_d$. As will
be seen, the correlation between the mixing angle $\theta$ and the
CP-odd neutral Higgs mass $M_A$ is important for cosmological study of
the model.

\bigskip\bigskip

\begin{figure}[t]
\begin{center}
\includegraphics[width=13cm]{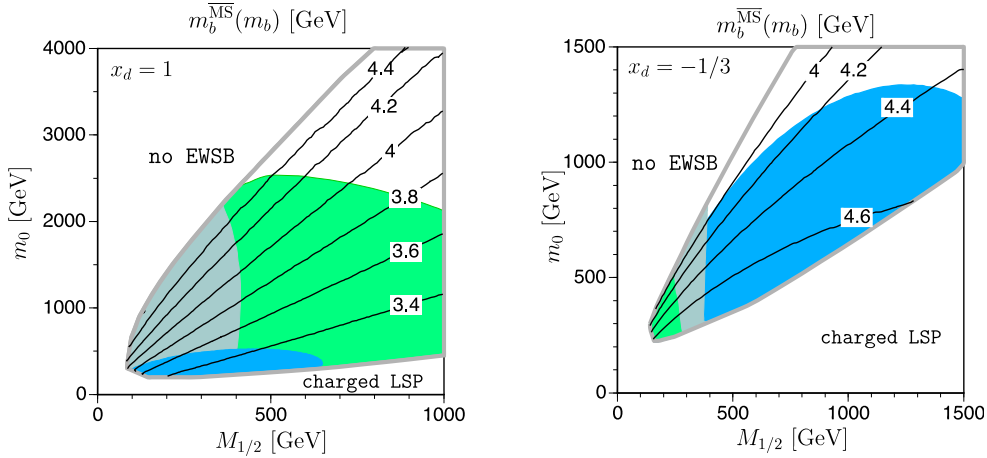}
\end{center}
\caption{The parameter space analysis for scalar and gaugino 
masses $m_0$ and $M_{1/2}$. The left (right) panel exhibits the result
for $x_d=1$, $\theta=60^\circ$ and $D/m_0^2=0.05$ ($x_d=-1/3$ 
,$\theta=55^\circ$ and $D/m_0^2=0.2$).
All the experimental mass bounds of superparticles are included. The
constraints from the lightest Higgs mass, $b\to s\gamma$ and 
$\tau\to\mu\gamma$ rare decays are shown as the gray, green and blue
shaded regions, respectively.\bigskip}
\label{fig:mbpre}
\end{figure}

The parameter space analysis is given in Fig.~\ref{fig:mbpre} for the
well-motivated limits of $x_d$: the aligned ($x_d=1$) and reduced 
($x_d=-1/3$) bottom/tau cases.
We simply set $A_0=0$ and show the bottom quark mass predictions in
the figures. The detail of trilinear couplings is irrelevant to our
qualitative discussion. The parameter analysis includes the
experimental lower bounds on superparticle masses~\cite{PDG}, the
flavor-changing decays of third-generation fermions 
($b\to s\gamma$~\cite{bsg} and $\tau\to\mu\gamma$~\cite{tmg}), and the
lightest Higgs mass bound. The general aspect of parameter space is
that, (i) if scalar superparticles are much heavier than gauginos, the
EWSB vacuum becomes unstable or the $\mu$ parameter is too small, and
(ii) the opposite side in the figure with light scalars is disfavored
in that the lightest superparticle is not charge neutral. Excepting
these trivially-excluded regions, in the green-shaded area in
Fig.~\ref{fig:mbpre} the prediction of $b\to s\gamma$ decay rate is
outside the experimental result, which we conservatively take 
as $2.0\times10^{-4}<{\cal B}(b\to s\gamma)<
4.5\times10^{-4}$~\cite{bsgexp}. In the blue-shaded region, 
the $\tau\to\mu\gamma$ branching ratio does not satisfy the upper
bound ${\cal B}(\tau\to\mu\gamma)<6.8\times10^{-8}$ 
(at 90\% C.L.)~\cite{taumugexp}. Finally the gray-shaded area leads to
the lightest Higgs mass being lower than the experimental lower bound
114.4~GeV~\cite{Hmassexp}, and hence excluded.

It is interesting to note that the constraints from flavor-violating
rare decays are rather different depending on the Yukawa 
parameters $\theta$ and $x_d$. In the aligned bottom/tau case with a
large $|x_d|$ (the left panel in Fig.~\ref{fig:mbpre}), 
the $b\to s\gamma$ constraint is important in a wide region of
parameter space with light superparticle spectrum. This is mainly due
to the negative sign of $\mu$ parameter imposed by the analysis of
bottom quark mass (see the right panel of Fig.~\ref{fig:mbtanb}). 
Since the standard model prediction of ${\cal B}(b\to s\gamma)$ is
consistent with the observed value, supersymmetric contribution should
be suppressed. For $\mu<0$ as in the usual Yukawa unification models
with universal supersymmetry-breaking parameters, relatively heavy
superparticles are required to suppress the amplitude, as seen in
Fig.~\ref{fig:mbpre} (the green-colored area in the left panel). On
the other hand, the reduced bottom/tau case with a 
small $|x_d|$ predicts a positive $\mu$, which sign makes a
cancellation among various diagrams operative, and 
the $b\to s\gamma$ constraint is not effective. 
In addition, a positive $\mu$ is suitable for a supersymmetric resolution
to the observed anomalous magnetic moment of the muon~\cite{KY}.
The lepton flavor violating decay $\tau\to\mu\gamma$ 
is however found to exclude a large
portion of parameter space in the reduced bottom/tau case (the
blue-colored area in the right panel). A source of flavor violation
comes from the D term which induces generation-dependent scalar lepton
masses (\ref{dl13}) and (\ref{dl2}). As we mentioned, a reduced
bottom/tau ratio allows a small $m_0$, leading to a large $|D|/m_0^2$,
and the $\tau\to\mu\gamma$ branching ratio is amplified.

\bigskip\bigskip

The distinctive types of superparticle spectra for the aligned and
reduced bottom/tau cases lead to different properties of the lightest
super particle (LSP). In the present model, LSP is the lightest
neutralino $\chi$ which is a linear combination of neutral 
gauginos $\tilde B$, $\tilde W$ and  
higgsinos $\tilde H_u$, $\tilde H_d$:
\begin{eqnarray}
  \chi \,=\, a_B\tilde B +a_W\tilde W +a_u\tilde H_u +a_d\tilde H_d.
\end{eqnarray}
We find that the LSP is a mixture of gauginos and higgsinos in the
allowed parameter region for the aligned bottom/tau case, while it
becomes almost a bino for the reduced bottom/tau case.

The LSP nature provides important information of the cold dark matter
in our universe. Assuming the R parity conservation as in the usual
supersymmetric standard models (to avoid too rapid nucleon decays),
the lightest neutralino $\chi$ is a suitable candidate for the cold
dark matter. The recent data brought out from the WMAP satellite gives
a strong limit on the relic density of the lightest neutralino as dark
matter~\cite{LSP_DM}. The central value of the dark matter density
extrapolated from the WMAP three years data 
is $\Omega_{\rm CDM}h^2=0.1045$~\cite{WMAP3}. It has been known that
a theoretical prediction of neutralino relic density $\Omega_{\chi}$
tends to much exceed $\Omega_{\rm CDM}$, in particular for a bino-like
LSP in the minimal supersymmetric standard model. In order to suppress
the predicted value of $\Omega_\chi$, the annihilation cross section
of the lightest neutralino must be enhanced. In our scenario, there
are two ways to have the enhancement: The first option is to increase
the amplitude $\chi\chi\to W^+W^-$ through the t-channel chargino
exchange, which requires a non-negligible component of higgsinos in
the LSP\@. The LSP inclusion of higgsino components needs a small size
of the $\mu$ parameter. The second one is to enhance the
annihilation into pairs of standard-model fermions through the
resonance near the CP-odd neutral Higgs boson. This can be achieved
with a tuned parameter region $2m_\chi\simeq M_A$.

\begin{figure}[t]
\begin{center}
\includegraphics[width=13cm]{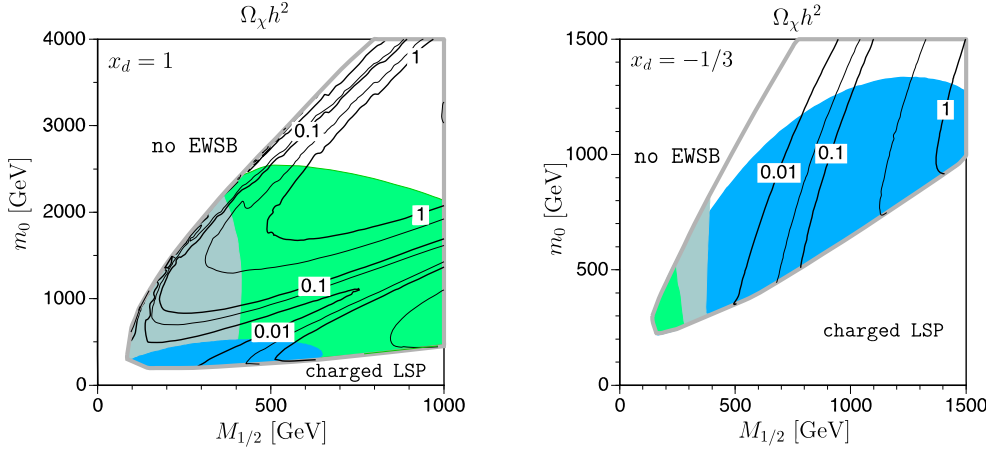}
\end{center}
\caption{The LSP neutralino relic density predicted from the
asymmetrical flavor structure. The experimental constraints 
and parameter fixing are taken in the same way as in 
Fig.~\ref{fig:mbpre}.\bigskip}
\label{fig:dm}
\end{figure}

Figs.~\ref{fig:dm} displays the predictions of neutralino relic
density $\Omega_\chi$ on the allowed parameter spaces described in
Fig.~\ref{fig:mbpre}.
For the aligned bottom/tau case with a large $|x_d|$ ($=1$ in the left
panel), both of the options are available: the relic density is
suppressed around the upper boundary of the correct EWSB region where
the $\mu$ parameter becomes small for $\Delta_b$ and the dark matter
somehow contains light higgsino components (we will see later what
amount of higgsinos is predicted). Another region is 
near $m_0\simeq M_{1/2}$, in which the LSP suppression is supplied by
the second option, that is, the CP-odd Higgs resonance. Around these
two regions, the cosmologically viable dark matter 
density $\Omega_{\rm CDM}h^2\simeq 0.1$ is explained by the LSP
neutralino. It is however noticed that the CP-odd Higgs boson
resonance is incompatible with other experimental data of the bottom
quark mass and/or the flavor-changing rare decay $b\to s\gamma$.

For the reduced bottom/tau case with a small $|x_d|$ ($=1/3$ in the
right panel of Fig.~\ref{fig:dm}), the neutralino relic density has
rather different behavior from the aligned bottom/tau case. We find
that only the second option, the CP-odd Higgs resonance, is available
to avoid the overproduction of the LSP neutralino. This is because, as
mentioned before, a small value of $|x_d|$ reduces the bottom/tau mass
ratio and requires a non-negligible (positive) threshold 
correction $\Delta_b$, which in turn imposes a not-so-small $\mu$
parameter producing non-negligible higgsino components in the lightest
neutralino. Therefore the resonant condition $2m_\chi\sim M_A$ should
be realized. It is remind of the EWSB fitting formula shown above
that, for some suitable value of the mixing parameter $\theta$, the
CP-odd neutral Higgs mass $M_A$ has a slight dependence of scalar soft
mass $m_0$. In this case, the gaugino mass $M_{1/2}$ determines $M_A$
and also the (almost bino) LSP mass. Therefore they are naturally
correlated to each other and the resonant condition is realized. This
fact can be seen in Fig.~\ref{fig:dm} (the right panel) that the relic
abundance of the LSP neutralino does not largely depend on the
universal scalar mass $m_0$. We checked that the prediction of CP-odd
neutral Higgs mass is also insensitive to $m_0$, and the 
relation $2m_\chi\sim M_A$ is approximately achieved in most of the
parameter space.

\bigskip\bigskip

Finally, to clarify the distinctive classes of mass spectrum and LSP
nature discussed above, we explore a wider parameter space by varying
the mixing parameter $\theta$ and the D term. The analysis is done for
fixed values of $x_d$ which is expected to be given by somewhat
discrete group-theoretical factor.

\begin{figure}[t]
\begin{center}
\includegraphics[width=14cm]{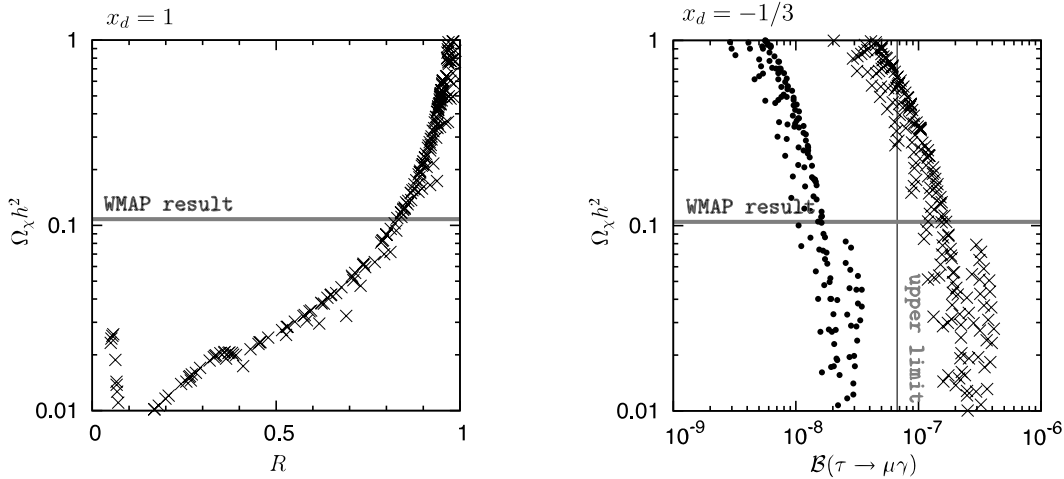}
\end{center}
\caption{The relic density of the LSP neutralino. The horizontal axis
in the left (right) figure denotes the gauge fraction in the lightest
neutralino (the $\tau\to\mu\gamma$ branching ratio). In the left
figure, we take $M_{1/2}=500$~GeV and $m_0=4$~TeV\@. In the right one,
we take $m_0=1$~TeV (dots) and $m_0=800$~GeV (crosses) 
with $M_{1/2}=500$~GeV.\bigskip}
\label{fig:scan}
\end{figure}

For the aligned bottom/tau case with a large $|x_d|$, heavier scalar
superparticles than gauginos are consistent with the observed bottom
quark mass and $b\to s\gamma$ branching ratio. The lightest neutralino
contains a sizable amount of higgsino components and the relic
abundance is suppressed. To see this, we introduce the gaugino
fraction of the lightest neutralino which is defined 
as $R_\chi\equiv|a_B|^2+|a_W|^2$. For a large value 
of $|x_d|\gtrsim 0.5$, a reduced $R_\chi$ is consistent with the
observed bottom quark mass. Fig.~\ref{fig:scan} (the left panel) shows
the prediction of LSP relic density for the aligned bottom/tau case
with $x_d=1$. In the figures, the experimental results for
superparticles, the $b\to s\gamma$ ratio, the masses of
third-generation fermions and Higgs bosons are taken into account. As
examples, the mass parameters are set to be $M_{1/2}=500$~GeV 
and $m_0=4$~TeV\@. The mixing parameter $\theta$ and D term are
scanned. The $\mu$ parameter is negative and heavy scalar
superparticles are compatible with the $b\to s\gamma$ constraint.
It is seen from the figure that there is a relation between the
gaugino fraction and the dark matter density. To explain the WMAP
data, the gaugino fraction is restricted as $R_\chi\lesssim 0.8$, that
is, the lightest neutralino is given by a mixture of bino and
higgsinos. This fact confirms the previous result that the abundance
of LSP neutralino is suppressed by the $\chi$ fusion.

The scanned parameters $\theta$ and D are limited by various
experimental results. For the aligned bottom/tau case with a large
value of $|x_d|$, $\theta$ is bounded from below by 
the $b\to s\gamma$ constraint as $55^\circ\lesssim\theta$. This is
converted to a moderate value of $\tan\beta$ from the argument 
of $\Delta_b$, which results in $38\lesssim\tan\beta\lesssim 48$. The
mixing parameter $\theta$ also has an upper bound,
$\theta\lesssim65^\circ$, from a cosmological reason by 
demanding $\Omega_\chi\leq\Omega_{\rm CDM}$, The allowed range of D
term is given by $0.02\lesssim D/m_0^2\lesssim 0.06$. A positive D
term decreases $|\mu|$ [see eq.~(\ref{mu})] and the upper bound of D
is given by the EWSB conditions or the mass bounds for the lightest
chargino and neutralino. As we have shown, heavy scalar superparticles
are required from $b\to s\gamma$. Therefore 
the $\tau\to\mu\gamma$ decay rate is sufficiently small and difficult
to be experimentally observed.

Another limit is the reduced bottom/tau case with $x_d=-1/3$. For such
a small value of $|x_d|$, superparticle masses tend to be degenerate
in low-energy regime. The CP-odd neutral Higgs mass is naturally
correlated to that of the lightest neutralino within an appropriate
range of $\theta$. The LSP is almost bino-like and the gaugino
fraction $R_\chi$ becomes almost unity. Fig.~\ref{fig:scan} (the right
panel) shows the prediction of LSP relic density as a function of 
the $\tau\to\mu\gamma$ branching ratio in the reduced bottom/tau case
with a typical example $x_d=-1/3$. All experimental constraints are
included while the parameters $\theta$ and D are scanned. We present
the results for two parameter examples: $m_0=800$~GeV 
and $M_{1/2}=500$~GeV (crosses) $m_0=1$~TeV and $M_{1/2}=500$~GeV
(dots). It is seen from the figure that the neutralino relic density
is suppressed around the dips where the annihilation amplitude is
enhanced through the CP-odd neutral Higgs 
resonance; $2m_\chi\simeq M_A$. The resonance condition is rather
independent on scalar masses, and therefore the flavor violation is
easily made within the current experimental bound.

The scanned regions of $\theta$ and D are limited in the
cosmologically allowed range. To keep the relation $2m_\chi\sim M_A$,
we find $\theta$ in the 
range $45^\circ\lesssim\theta\lesssim62^\circ$. This corresponds 
to $43\lesssim\tan\beta\lesssim53$, which is larger than the aligned
bottom/tau case. The D term is constrained to a 
range $0.03\lesssim D/m_0^2\lesssim0.4$ by the observed bottom quark
mass and the requirement for dark matter being charge neutral. Thus in
the reduced bottom/tau case, $\tan\beta$ and D become larger than the
aligned bottom/tau case. That is a general property 
for $|x_d|\lesssim0.5$.

It is interesting to note that flavor-changing transitions are
important in the reduced bottom/tau case. This is due to the following
two reasons. The first is the light mass spectrum of scalar
superparticles than that in the aligned bottom/tau case. Another is the
generation-dependent D-term contribution to scalar masses, which leads
to the non-degeneracy in the down/charged lepton sectors (\ref{dl13})
and (\ref{dl2}). As seen before, a smaller $|x_d|$ (the reduced
bottom/tau case) predicts a larger value of D term (and a 
larger $\tan\beta$), and enhances the flavor-violating decay
amplitudes. The $\tau\to\mu\gamma$ branching ratio is shown in the
right panel of Fig.~\ref{fig:scan}. The current experimental upper 
limit ${\cal B}(\tau\to\mu\gamma)<6.8\times10^{-8}$ gives a lower
bound of superparticle masses. For $m_0\simeq 1$~TeV, the decay rate
becomes smaller than the current upper limit and is within a range of
near future experiments.

\medskip

As a final remark, we consider the dileptonic rare decay of B 
meson $B^0\to\mu^+\mu^-$ recently studied. For a large value 
of $\tan\beta$, the branching ratio approximately scales 
as $\tan^6\beta/M_A^4$~\cite{Bmumu}. The experimental upper bound of
the ratio gives a constraint for model parameters unless rather
heavy CP-odd neutral Higgs boson is assumed. In our 
scenario, $\tan\beta$ is relatively large and the predicted branching ratio
is expected to be. In particular, for the reduced bottom/tau case,
the CP-odd neutral Higgs boson is allowed to have a similar mass to
the LSP for suppressing the neutralino relic density. As a result, the
gaugino mass parameter $M_{1/2}$ would be rather restricted. For the
aligned bottom/tau case, the $b\to s\gamma$ constraint requires
TeV-scale CP-odd neutral Higgs boson and the constraint from
dileptonic decay is not severe.

\bigskip\bigskip

In summary, we have studied the supersymmetric grand unified theory
with the Yukawa coupling form (\ref{Y}). This flavor structure is
suggested by the recent experimental data of fermion masses and mixing
angles. The two parameters $\theta$ and $x_d$ in the flavor sector are
important and control the low-energy property of the model. These
parameters are related to supersymmetry-breaking threshold corrections
to the bottom quark mass and the successful electroweak symmetry
breaking. In this letter we have investigated the flavor-changing rare
decays of heavy fermions and also the LSP neutralino relic as dark
matter of the universe. For the aligned bottom/tau case 
with $x_d\sim{\cal O}(1)$, a hierarchical superparticle spectrum is
obtained; heavier scalars than gauginos and higgsinos. The lightest
neutralino contains a significant amount of higgsino components. We
have found that the reduced bottom/tau case with a small $|x_d|$ is
particularly interesting to low-energy phenomenology: (i) the
superparticles are light and degenerate in masses, like a recent
analysis in~\cite{MM}. (ii) the $\mu$ parameter becomes positive,
which provides a simultaneous supersymmetric solution to the
experimental data of bottom/tau mass ratio, the $b\to s\gamma$ rare
decay, and the muon anomalous magnetic moment. (iii) a gaugino-like
LSP is expected and the CP-odd neutral Higgs boson is as light as the
LSP for explaining the dark matter density of the universe, and (iv)
finally the D term is significant for the dark matter
and the lepton flavor violation.
Its contribution to scalar masses is generation
dependent due to the lopsided flavor structure. These
features induces distinct and detectable signatures in low-energy
phenomenology and future searches for supersymmetry will probe the
validity of the scenario.

\bigskip
\subsection*{Acknowledgments}
This work is supported in part by the scientific grant from the
Ministry of Education, Science, Sports, and Culture of Japan
(No.~17740150) and by grant-in-aid for the scientific research on
the priority area \#441 (No.~16081209).

\end{document}